\documentclass[twocolumn]{aastex631}

\graphicspath{{./}{figures/}}

\shorttitle{Asteroids and life}
\shortauthors{R. G. Martin \& M. Livio}

\begin{document}

\title{Asteroids and life: How special is the solar system?}

\author{Rebecca G. Martin}
\author{Mario Livio}
\affil{Department of Physics and Astronomy, University of Nevada, Las Vegas,
4505 South Maryland Parkway, Las Vegas, NV 89154, USA}





\begin{abstract}
Asteroid impacts with the Earth may have played an essential role in the emergence of life on  Earth through their creation of favorable niches for life, changes to the atmosphere and delivery of water. Consequently, we suggest two potential requirements for life in an exoplanetary system: first, that the system has an asteroid belt, and second, that there is a mechanism to drive asteroids to impact the terrestrial habitable planet. Since in the solar system, the $\nu_6$ secular resonance has been shown to have been important in driving these impacts,   we explore how the masses and locations of  two giant planets determine the location and strength of this secular resonance. Examining observed exoplanetary systems with two giant planets, we find that a secular resonance within the asteroid belt region may not be uncommon. Hence the solar system is somewhat special, but the degree of fine-tuning that may be necessary for the emergence of life is not excessive. Finally, with $n$-body simulations, we show that when the two giant planets are close to the 2:1 mean motion resonance, the asteroid belt is unstable but this does not lead to increased asteroid delivery. 
\end{abstract}



\section{Introduction}

Studies of the origin of life on Earth have made an impressive progress in the last decade  \citep[see, e.g.][for recent reviews]{Szostak2017,Szostak2017b,Sutherland2016,Sutherland2017}.  While many questions remain open, there is a growing consensus on many aspects of the processes involved. For example, it appears that hydrogen cyanide, which today is considered a deadly poison, may have provided the primordial pathway from prebiotic chemistry to life.  In this work we are interested specifically in the role that asteroid impacts may have played in the emergence of life on Earth. Unfortunately, one of the effects of plate tectonics has been to erase most of the traces of the early geological evolution of the Earth’s surface. However, most researchers agree that the Earth had experienced a relatively high impact rate during roughly the first billion years of its existence \citep[e.g][]{Bottke2017}.
 
 
For a few decades, the dominant opinion has been that asteroid impacts have impeded rather than aided the emergence of life on Earth \citep[see, e.g., discussion by][]{Sleep2018}. More recently, however, opinions started to swing in the other direction, suggesting that asteroid impacts may have, in fact, been essential for the transition from chemistry to biology \citep[see, e.g.,][for an extensive review and references therein]{Osinksi2020}. According to the new scenario, impacts may have acted both as generators of favorable niches for life to emerge (niches such as impact crater lakes and shocked rocks), and as the agents which had changed the entire Earth environment in a way that made it conducive for life (e.g. by producing atmospheric hydrogen cyanide). Asteroid impacts may have delivered water to the surface of the Earth \citep{Morbidellietal2000,Martin2021water} and led to the formation of the Moon \citep[e.g.][]{Cuk2012S,Canup2012}.


Another crucial effect for the origin of life could have been created by relatively large impacts during the “late veneer” (material accreted by Earth after the formation of the Moon). Such impacts could have produced a reduced atmosphere of the early Earth, when the iron core of the impactor reacted with water in the oceans. As the iron oxidized, the hydrogen was released, resulting in an atmosphere (which could have lasted a few million years) favoring the emergence of simple organic molecules  (e.g., \citealp{Zahnle2020}; see also \citealp{Genda2017a,Genda2017b,Benner2019}).

Given the mounting evidence for the potential role of asteroids in the emergence of life on Earth, it is not unreasonable to {\it assume} that asteroid impacts on a terrestrial planet (in the habitable zone of its host star) are a necessary condition for the emergence of life, and to study the consequences of this assumption. This hypothesis dictates two requirements for an exoplanetary system, for it to harbor life: 
\begin{enumerate}
    \item The system has  to contain the equivalent of an asteroid belt;
    \item The system has to have a mechanism which drives asteroids out of the belt and causes them to impact the terrestrial planet. 
\end{enumerate}
In the present work we explore under which conditions planetary systems can satisfy these two requirements, and the implications of these conditions for the important question of whether the solar system is in any way special, when compared to other confirmed exoplanetary systems. In Section 2 we examine the formation of asteroid belts. In Section 3 we explore the location and strength of secular resonances with different giant planetary system architectures and compare to observed exoplanetary systems.   In Section 4 we run simulations to test the efficiency of asteroid impacts from the belt.
A discussion and conclusions follow in Section~5.



\section{The formation of asteroid belts}

 The snow line radius is the distance from the central star outside of which water is found in the form of ice. It occurs at a temperature of around $T_{\rm snow}=170\,\rm K$ in the protoplanetary disk \citep{Hayashi1981,Lecaretal2006}. The snow line radius evolved with the protoplanetary gas disk, likely beginning farther out and moving inwards as the disk temperature cooled \citep{GaraudandLin2007,MartinandLivio2013snow}.  In our solar system  today, the composition of the asteroids in the belt has a transition at a radius of about $2.7\,\rm au$  \citep{DeMeo2014} suggesting that the snow line radius may have been located there in the asteroid belt at the time that the gas disk dissipated \citep{Abeetal2000,Morbidelli2010}.  Recent studies have shown that the compositional gradient may be explained by the gas giant’s growth and/or migration \citep[e.g.][]{Raymond2017}. 

In protoplanetary disks, the solid mass density increases significantly outside of the snow line \citep{Pollack1996}. Consequently giant planets likely form outside of the snow line radius \citep{Kennedy2008} and
asteroid belts likely form around the location of the water snow line radius, inside of the giant planets \citep{MartinandLivio2013asteroids}. Observations of debris disks show that it is common for there to be two components to the disk, similar to the asteroid belt and the Kuiper belt in the solar system \citep{Kennedy2014,Geiler2017,Rebollido2018}. The gap between the two belts is likely caused by planet formation. \cite{Ballering2017} found that the warm components in single-component systems follow the primordial snow line, so they likely arise from asteroid belts \citep[see also][]{Morales2011}.

Observations of exoplanets initially suggested that Jupiter was somewhat of an outlier in terms of its orbital location \citep{Beer2004,Martin2015}. However, more recent exoplanet detections with the radial velocity  and microlensing methods have shown that the location of Jupiter is not particularly special. In fact, there is a peak in the occurrence of giant planets at around $2-3\,\rm au$, close to the snow line radius \citep{Fernandes2019,Nielsen2019}. However, the frequency of planets with masses of $0.1-20\,M_{\rm J}$ (where $M_{\rm J}$ is the mass of Jupiter) between 0.1 and 100$\,\rm au$ is only about $26\%$ and it drops to $6.2\%$ if only planets more massive than Jupiter are included. Thus, Jupiter-like planets are somewhat rare \citep[see also][]{Wittenmyer2016}. Note that giant exoplanets that are found close to their host stars have likely undergone some type of migration from where they formed.

In this work we consider the classical picture in which an asteroid belt forms because perturbations from an outer giant planet prevent material from forming a planet.  Violent collisions  lead to fragmentation rather than accretion \citep{Wetherill1989}.\footnote{Several new studies suggest that the belt in the solar system may have initially been empty and later populated by planetesimals scattered from other locations \citep{Raymond2017b,Raymond2020}. In this scenario, a giant planet may not be a requirement for an asteroid belt to form. } For a relatively small giant planet eccentricity, the outer edge of an asteroid belt likely occurs roughly at the location of the 2:1 mean motion resonance (MMR) with the giant planet. In the solar system this is at about $a_{\rm out}=3.3\,{\rm au} = 0.63\,a_{\rm J}$, where $a_{\rm J}$ is the semi-major axis of Jupiter. The inner edge of the asteroid belt is not so well defined. It should be at the radius at which the perturbations from the giant planet become small enough so that planet formation can proceed. Based on the solar system, we take in this work the inner edge to be close to the orbit of Mars, $a_{\rm in}=1.5\,{\rm au}= 0.29\,a_{\rm J}$. We also assume that the radial extent of an asteroid belt scales simply with the location of the innermost giant planet.

\begin{figure*}
\begin{center}
\includegraphics[width=0.7\columnwidth]{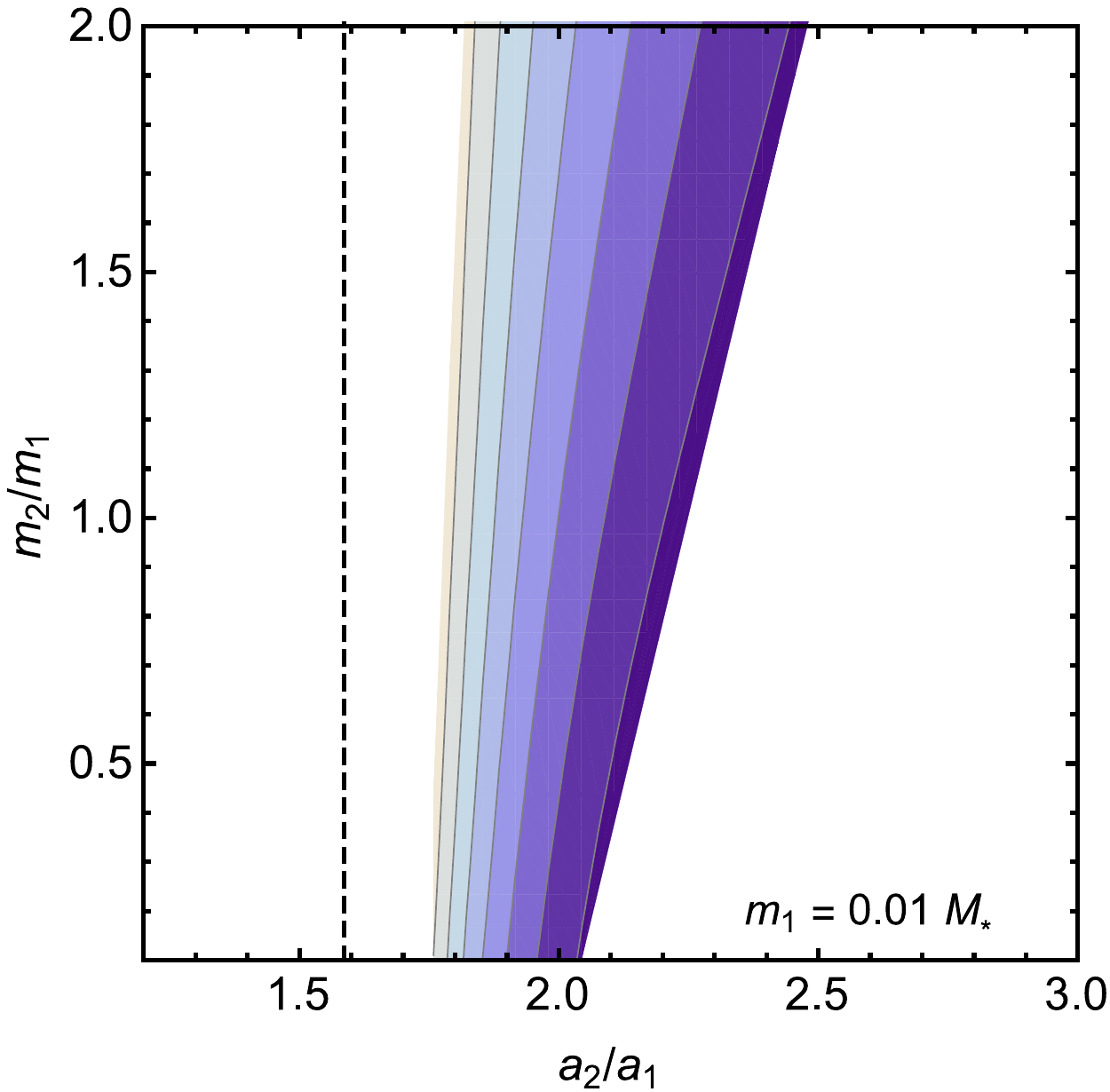} 
\includegraphics[width=0.84\columnwidth]{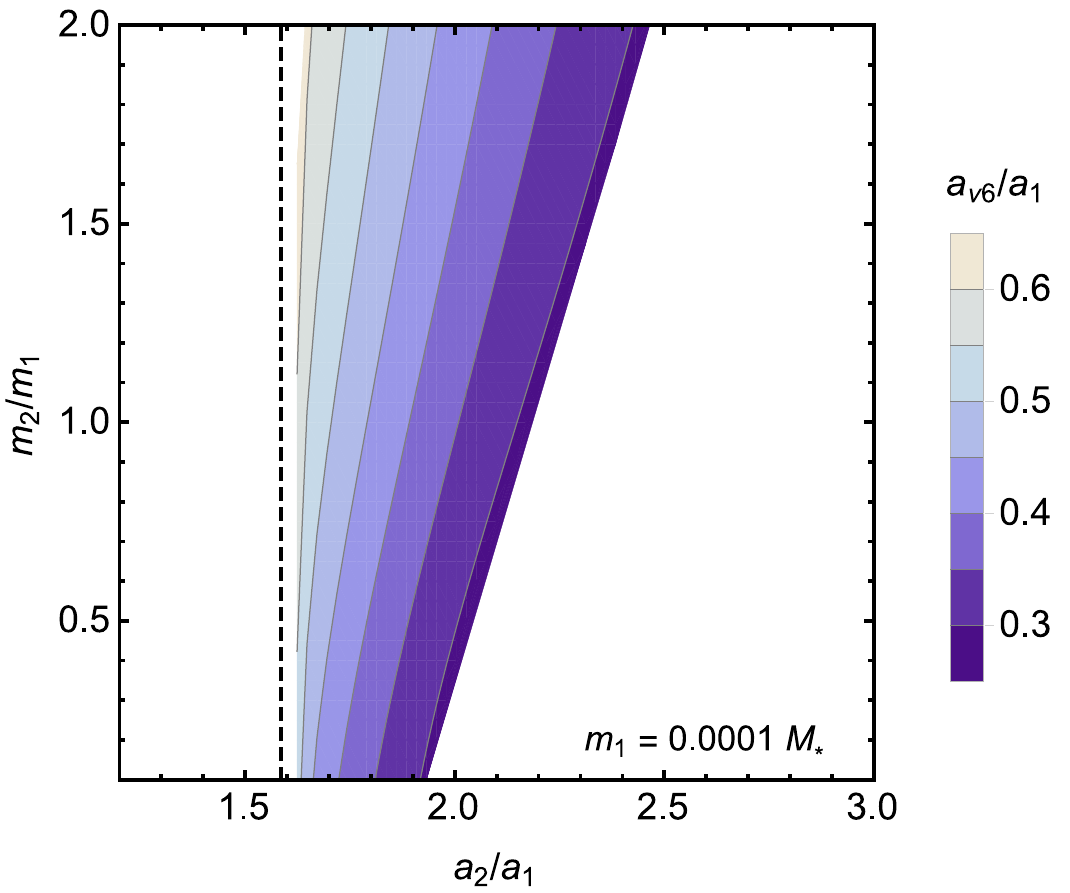} 
\hspace*{-1.2cm}
\includegraphics[width=0.7\columnwidth]{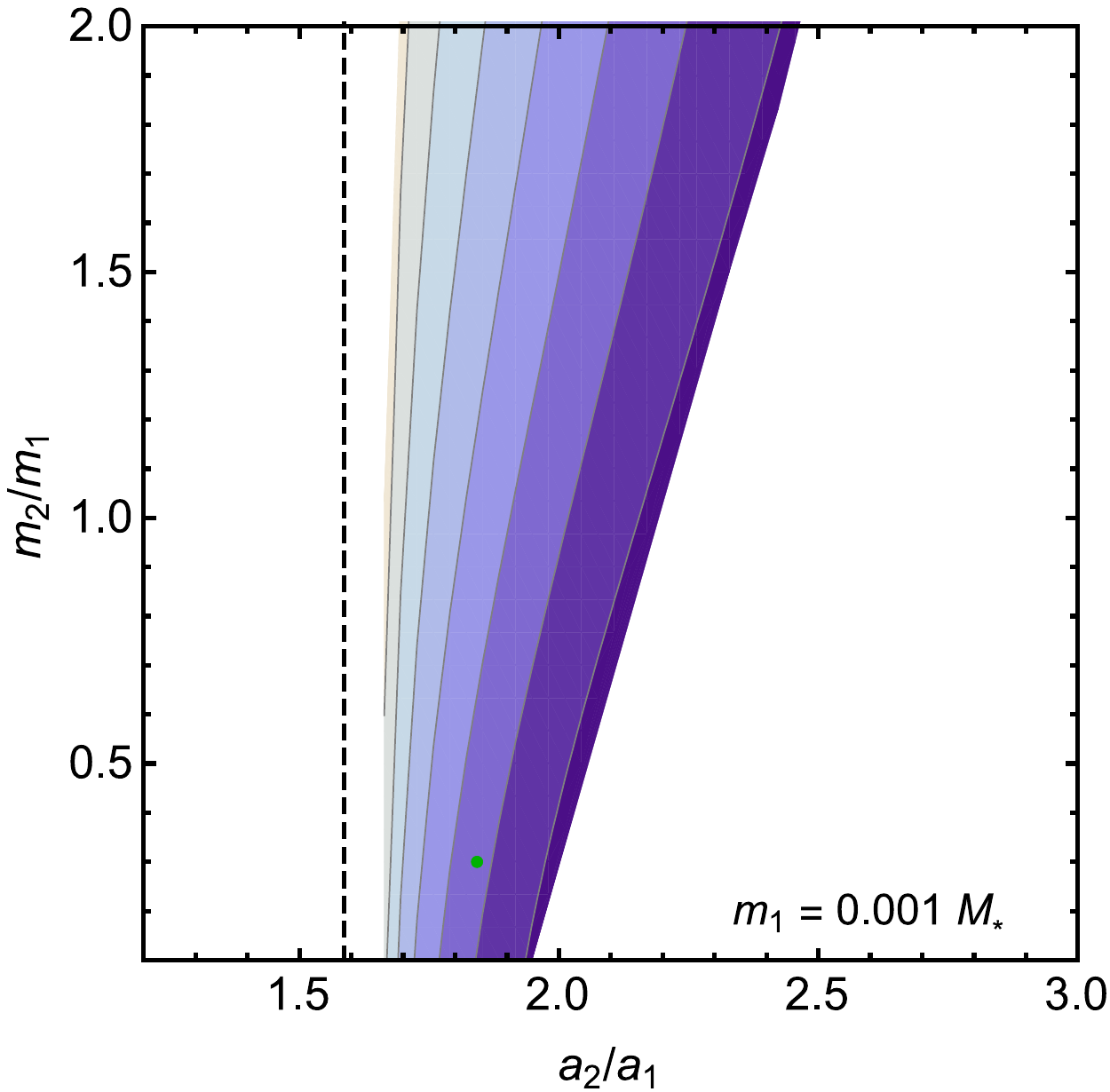} 
\includegraphics[width=0.7\columnwidth]{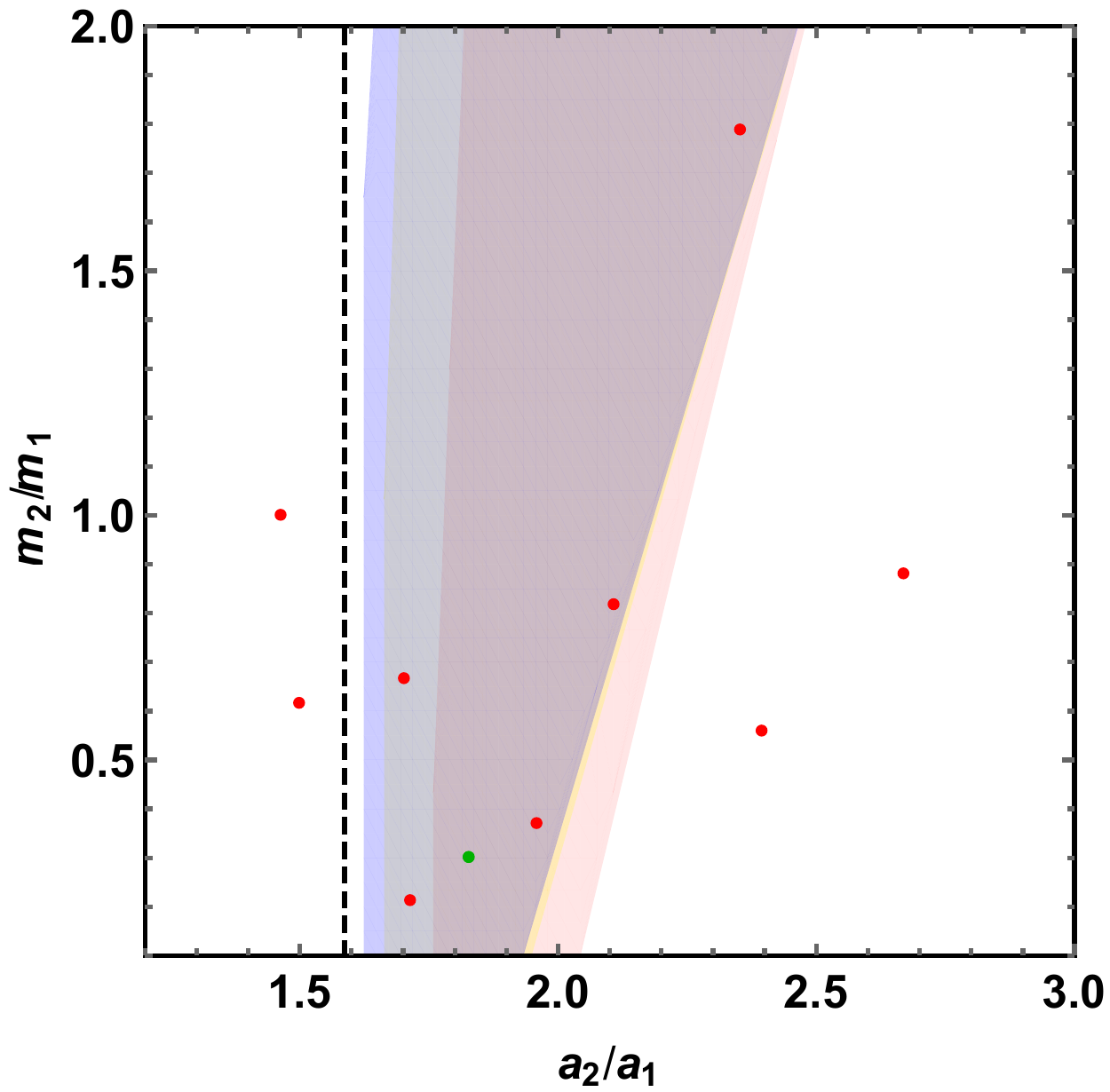}  
	\end{center}
    \caption{Location of the resonance  as a fraction of the inner giant's semi-major axis, $a_{\nu_6}/a_1$. The colored regions show where the resonance is  in the radial range $0.29<a_{\nu_6}/a_1<0.63$ for varying mass and semi-major axis of the outer giant planet scaled to those of the inner planet. The dashed vertical lines represent the 2:1 MMR between the two giants. The inner planet has a mass of $m_1=0.01\,M_\star$ (top left), $m_1=0.001\,M_\star$ (bottom left) and $m_1=0.0001\,M_\star$ (top right). The green points shows the solar system. Bottom right: Relative mass and semi-major axis of exoplanet planet pairs shown in Table~\ref{table2} (red points). The shaded regions show the where the $\nu_6$ resonance lies within the asteroid belt region for inner planet masses $m_1=0.0001\,M_\star$ (blue), $m_1=0.001\,M_\star$ (yellow) and $m_1=0.01\,M_\star$ (red). 
}
    \label{contour}
\end{figure*}

\section{The $\nu_6$ secular resonance}

\begin{figure*}
\begin{center}
\includegraphics[width=0.7\columnwidth]{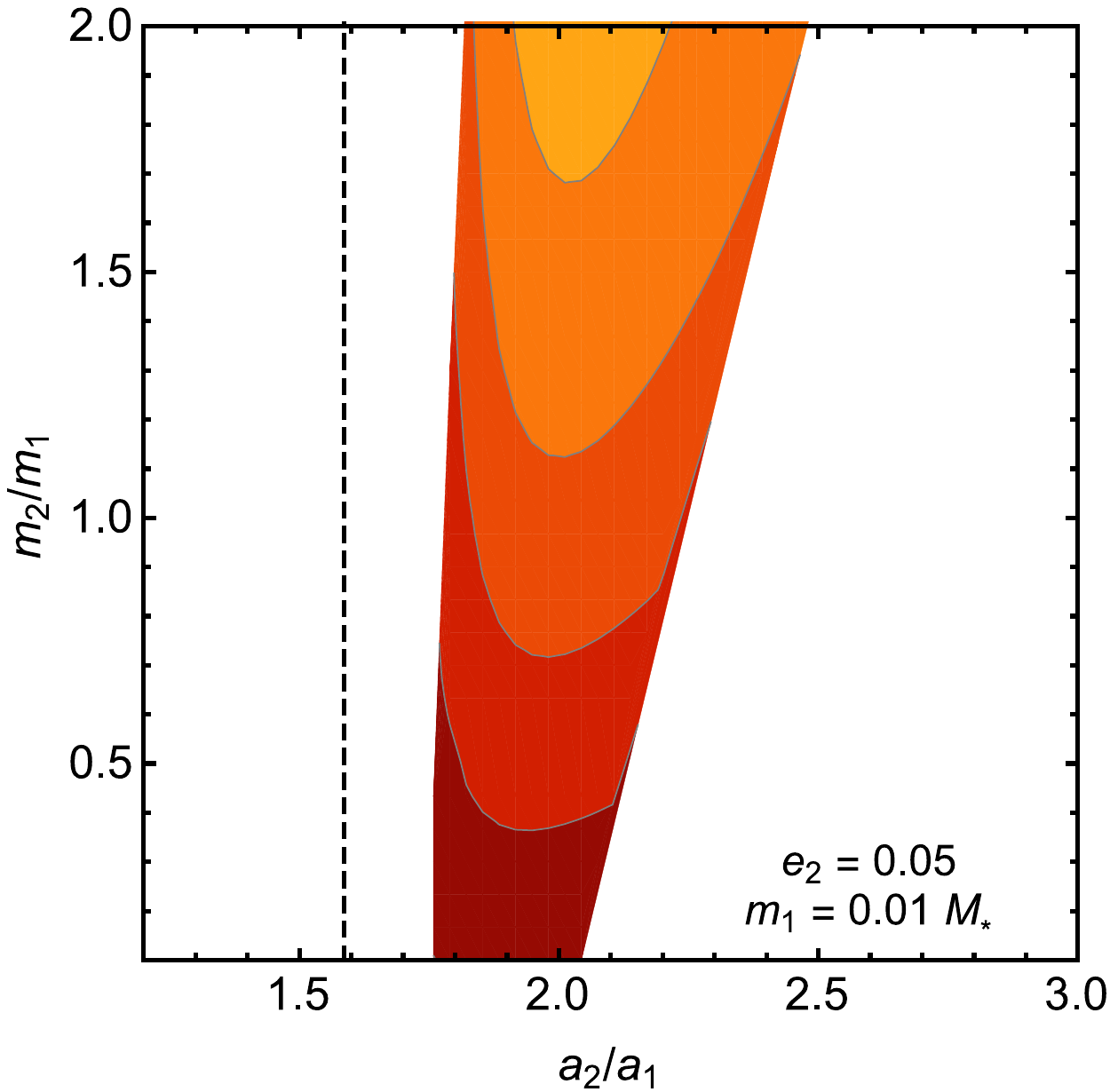} 
\includegraphics[width=0.84\columnwidth]{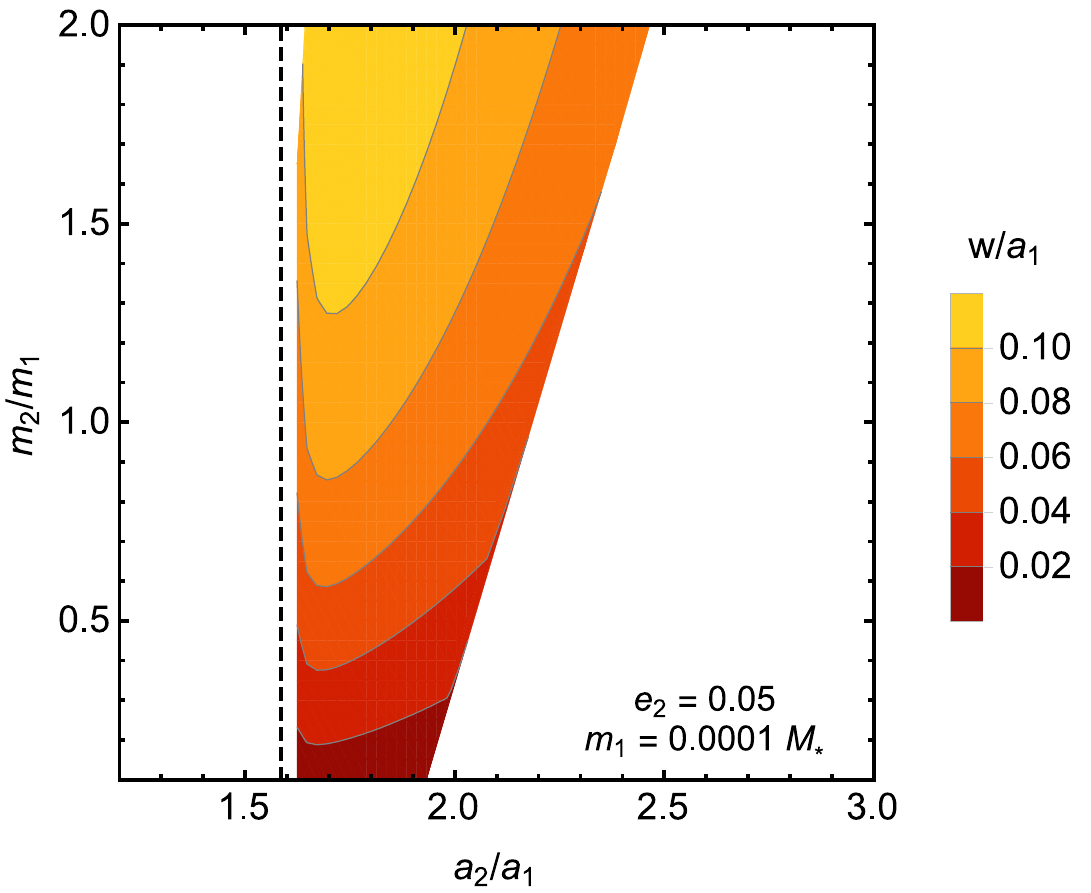} 
\hspace*{-1.2cm}
\includegraphics[width=0.7\columnwidth]{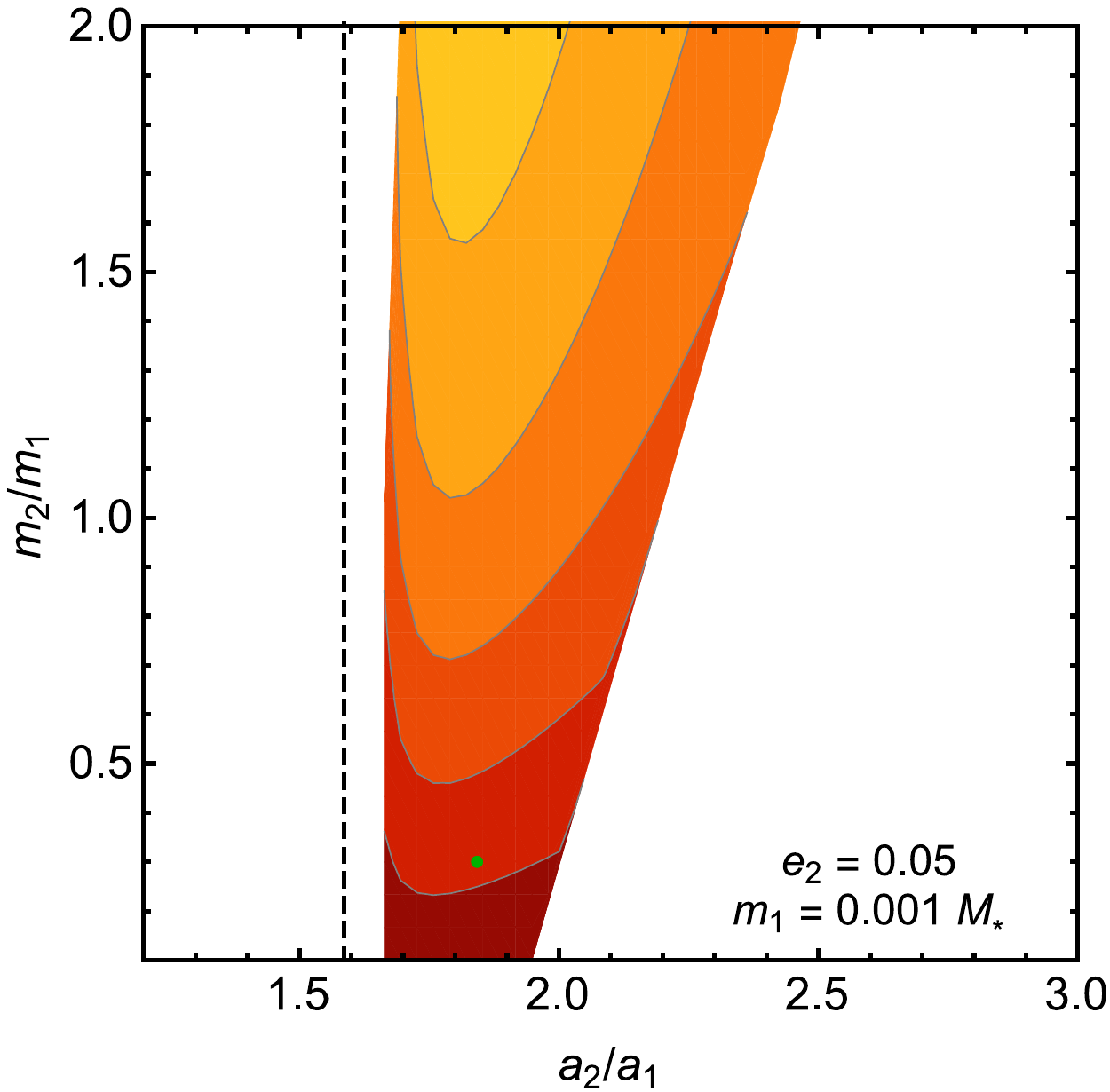} 
\includegraphics[width=0.7\columnwidth]{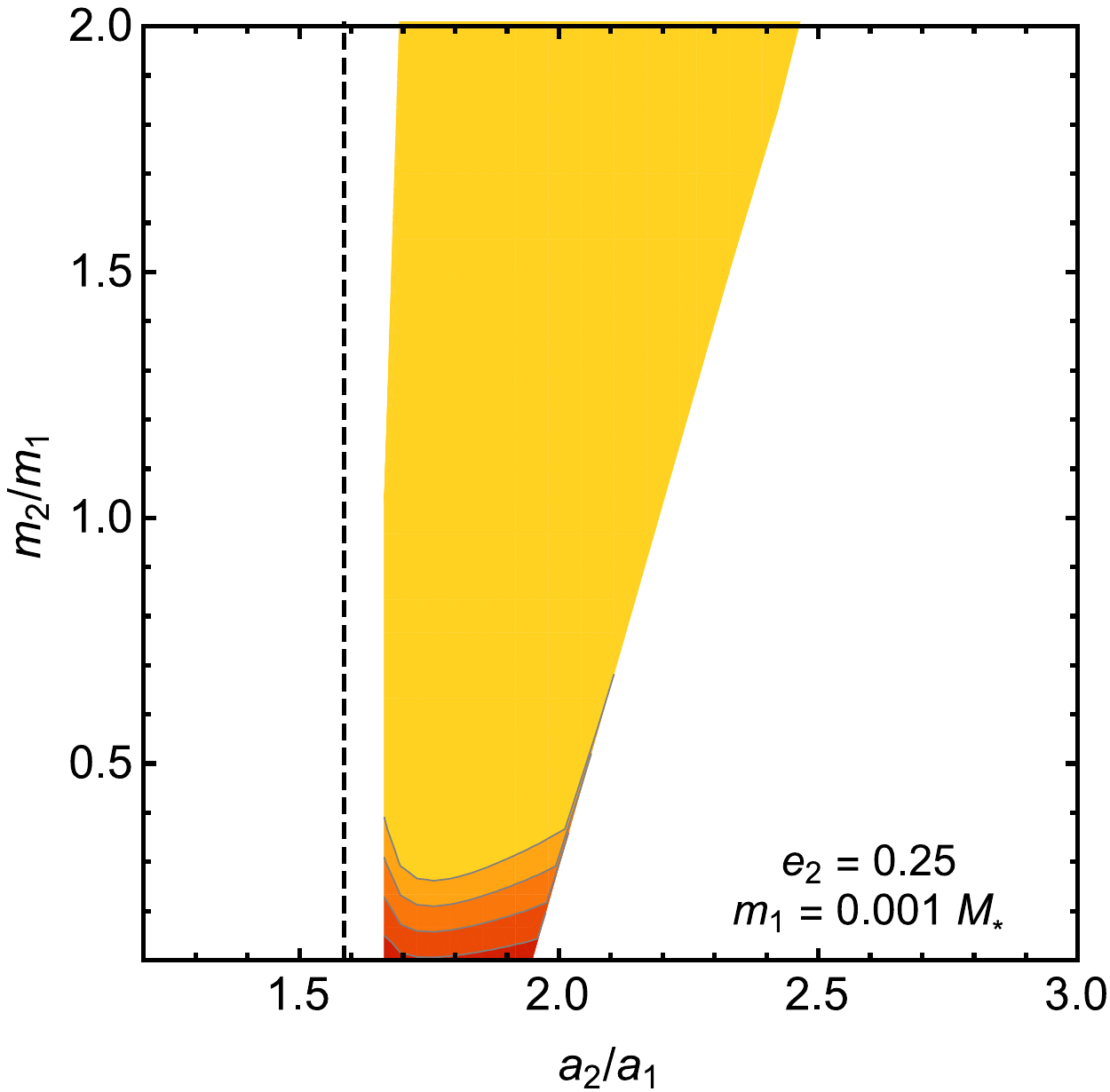} 
	\end{center}
    \caption{ The width of the resonance, $w/a_1$, defined to be where $e_{\rm f}>0.08$. The inner planet has a mass of $m_1=0.01\,M_\star$ (top left), $m_1=0.001\,M_\star$ (bottom left) and $m_1=0.0001\,M_\star$ (top right) and the outer planet has eccentricity $e_2=0.05$. The lower right panel shows $m_1=0.001\,M_\star$ and $e_2=0.25$. The inner planet eccentricity is $e_1=0.05$ in all cases.} 
    \label{contour2}
\end{figure*}

Most solar system models to date have focused on the idea that the giant planets cleared out material from the asteroid region \citep{Petit2001,Chambers2001b,OBrien2007,Nesvorny2021}.\footnote{Note that planetary embryos that are embedded in an asteroid belt may lead to perturbations and eccentricity excitation although models all include Jupiter and Saturn at their current location and hence also involve resonances \citep[e.g.][]{OBrien2007}.} While MMRs may play a role in asteroid delivery, in the solar system, the $\nu_6$ resonance is the strongest resonance in the asteroid belt \citep[e.g.][]{Froeschle1986,Morbidelli1991}. The resonance occurs where the precession rate of a test particle is equal to the eigenfrequency of Saturn. It causes asteroids in the region to become eccentric and collide with the Earth with a relatively high probability of a few percent \citep{Ito2006,Haghighipour2016,Martin2021water}. Without the $\nu_6$ resonance in the asteroid belt, the number of asteroid collisions with the Earth is significantly reduced \citep{Morbidelli1994, Bottke2000,Smallwood2018}. 

\subsection{Location of the $\nu_6$ resonance}

We examine the location of the $\nu_6$ resonance for varying giant planet architectures with a linear theory \citep{Dermott1986,Murray2000}, including a correction due to the 2:1 MMR between the planets \citep{Malhotra1989,Minton2011}. 
The giant planets were taken to have masses of $m_1$ and $m_2$ and orbits with semi-major axes $a_1$ and $a_2$, respectively, around a host star of mass $M_\star$. We vary the relative semi-major axis and mass of the outer giant planet. The eccentricities of the planets were assumed to be $0.05$ unless otherwise stated.  

Fig.~\ref{contour} show the location of the $\nu_6$ resonance for three different inner planet masses.  The colored regions show where the resonance falls in the approximate range for an asteroid belt given by $0.29<a_{\nu_6}/a_1<0.63$. The vertical dashed line shows the location of the 2:1 resonance between the two planets.  
The location of the $\nu_6$ resonance  does not change significantly with the mass ratio of the planets but it is very sensitive to the semi-major axis ratio. Generally, the resonance moves outwards with decreasing semi-major axis of the outer planet, until the planets approach the 2:1 MMR and there is no longer a $\nu_6$ resonance within the belt.   Note that had we chosen a larger inner radius for the belt, the colored region would be truncated on the right hand side and the planets would need to be closer together in order for a resonance to exist in the belt.

We should note that a secular resonance could also be formed with a giant planet and an outer binary star companion. While the star is much more massive than a giant planet, it is also much farther away \citep[see also][]{Smallwood2018b}, but in this work we focus on the effects of planetary systems.

\subsection{Strength of the $\nu_6$ resonance}

While the location of the resonance doesn't vary much with outer planet mass, the strength of the resonance does. 
Fig.~\ref{contour2} show the width of the secular resonance, $w$, which we define  to be the radial range where the forced eccentricity of a test particle is $e_{\rm f}>0.08$. We only show the width for parameters for which the resonance lies within the belt.  Generally, the closer in and higher the mass of the  outer planet (for fixed inner planet mass and semi-major axis), the wider the resonance and therefore the higher the number of asteroids that collide with the terrestrial planet. The eccentricities of the giant planets do not affect the location of the resonance but they do affect the width. 
A higher eccentricity of the inner planet leads to an increase in the average eccentricity of the belt but does not significantly change the resonance. However, the outer planet eccentricity has a strong effect on the resonance. The lower right panel in Fig.~\ref{contour2} shows a higher eccentricity outer planet. The resonance may be very wide and destroy the entire belt on a short timescale for a high eccentricity outer planet. In this case there may be initially a very high rate of impacts but a much lower rate on long timescales.

\subsection{The $\nu_6$ resonance in exoplanetary systems}

\begin{table*}
\begin{center}
\begin{tabular}{l l l l l l l l l l l ccc}
\hline
 Host star  & $M_\star/\rm M_\odot$ &  $N_{\rm p}$ & P1 & P2 & $a_1/\rm au$ & $a_2/\rm au$ & $m_1/\rm M_{\rm J}$ & $m_2/\rm M_{\rm J}$  & $a_{\nu_6}/a_1$ & $e_1$ & $e_2$ & $w/a_1$\\ 
 \hline
 \hline 
 The Sun & 1.00 & 8 & J & S & 5.20 & 9.58 & 1.00 & 0.30  & {\bf 0.37} & 0.049 & 0.052 &  0.025  \\
HD 34445* & 1.07 & 6 &  b &  g & 2.07 & 6.36 & 0.82 & 0.38  & 0.10 & & &  \\
HD 141399 & 1.07 &  4 & d &  e & 2.09 & 5.00 & 1.18 & 0.66  & 0.20 & & &\\
47 Uma & 1.06 & 3 &  b &  c & 2.10 & 3.60 & 2.53 & 0.54  & {\bf 0.53} & 0.032 & 0.098 &0.021 \\
OGLE-2006-BLG-109L & 0.51 & 2 &  b &  c & 2.30 & 4.50 & 0.73 & 0.27 & {\bf 0.32} & - & 0.15 &0.066 \\
UZ For & 0.70 & 2 &  c &  b & 2.80 & 5.90 & 7.70 & 6.30  &{\bf 0.33} & 0.05 & 0.04 &0.037 \\
NN Ser* & 0.54 & 2 & d &  c & 3.43 & 5.35 & 2.30 & 7.33  & 0.92 && & \\
HD 66428* & 1.05 & 2 & b &  c & 3.47 & 23.0 & 3.20 & 27.0  & 0.094 & & &\\
HU Aqr AB & 0.88 & 2 &  b &  c & 3.60 & 5.40 & 6.00 & 4.00  & 0.80 & & &\\
HD 30177 & 0.99 & 2 & b &  c & 3.70 & 9.89 & 8.62 & 7.60 & 0.17 & & &\\
HD 50499 & 1.31 & 2 &  b &  c & 3.83 & 9.02 & 1.64 & 2.93  & {\bf 0.30} &0.266 & 0.0 & 0.10\\
OGLE-2012-BLG-0026L* & 1.06 & 2 &  b &  c & 4.00 & 4.80 & 0.15 & 0.86  & 0.90 & && \\
HR 8799 & 1.61 & 4 & e &  d & 16.4 & 24.0 & 10.0 & 10.0 &  0.77 & & &\\ 
PDS 70 & 0.76 & 2 &  b &  c & 20.0 & 34.0 & 3.00 & 2.00  & 0.65 & & &\\
\hline
\end{tabular}
\end{center}
\caption{Exoplanetary systems with two or more planets with mass greater than $0.1\,M_{\rm J}$ with semi-major axis greater than $2\,\rm au$. Column~1 and~2 show the name and mass of the host star. Column~3 shows the number of detected planets in the system. Columns~4-9 show the names, semi-major axes and masses of the giant planets. Column~10  shows the location  of the $\nu_6$ resonance. If the resonance is within the belt, the location is shown in bold and then in Columns 11, 12 and 13 we show the planet eccentricities and the width of the resonance. Where the eccentricity is unknown, we take $e_1=0.05$ to calculate the width. The width is where the forced eccentriticy $e_{\rm f}>0.08$ except for HD~50499 where we take $e_{\rm f}>0..3$. *System is out of the range of Fig.~\ref{contour}.}
\label{table2}
\end{table*}

Table~\ref{table2} shows all of the multi-planet systems in the NASA Exoplanet Archive for which at least two planets have both their semi-major axis and mass determined. We have kept only planets that are at an orbital semi-major axis greater than $2\,\rm au$ and which have a mass greater than $0.1\,\rm M_{\rm J}$. We removed the system TYC 8998-760-1 since its planets are at distances of hundreds of au.  For systems that have more than two giant planets that satisfy our criteria, we considered the innermost two planets. For each planet pair we estimated the location and width of the $\nu_6$ resonance. In the calculations we only included the two planets represented in the Table.  The red points in  the lower right panel in Fig.~\ref{contour} show the observed data.  We also shaded three regions that show where the location of the $\nu_6$ resonance falls within the asteroid belt region for different inner planet masses.

Due to the difficulty in detecting giant planets at large orbital radii, the observed systems are significantly affected by selection effects. However, out of the thirteen systems (other than the solar system), we find that four have a configuration such that a secular resonance would fall in an asteroid belt (if it exists).  This suggests that the particular requirement to have  a secular resonance falling within the asteroid belt region does not in itself imply a very stringent fine-tuning on exoplanetary systems for them to allow the emergence of life.

\section{Simulations of asteroid impacts}
\label{simulations}

\begin{figure*}
\begin{center}
\includegraphics[width=0.66\columnwidth]{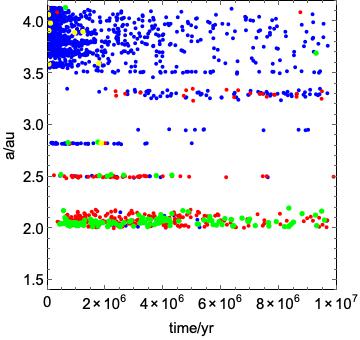}
\includegraphics[width=0.66\columnwidth]{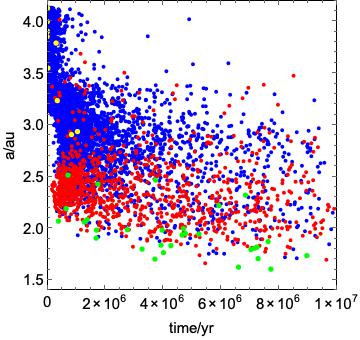}
\includegraphics[width=0.66\columnwidth]{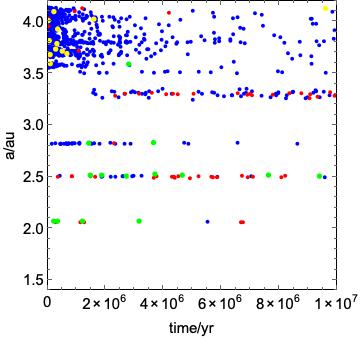} 
	\end{center}
    \caption{Asteroid outcomes as as a function of time and their original semi-major axis, $a$, for run1 (left), run2 (middle) and run3 (right). Blue points show asteroids that are ejected. Red points show asteroids that collide with the Sun, green points show asteroids that collide with the Earth and yellow points show asteroids that collide with Jupiter or Saturn.  } 
    \label{splash}
\end{figure*}

\begin{table*}
\begin{tabular}{l l l l l l l l l l}
\hline
 Simulation   & $a_{\rm J}/\rm au$ & $a_{\rm S}/\rm au$ & $N_{\rm Sun}$ & $N_{\rm Earth}$ & $N_{\rm J,S}$ &  $N_{\rm eject}$ &  $N_{\rm outcome}$ & $N_{\rm remain}$ & $P_{\rm collide}$   \\
 \hline
 \hline
run1 & 5.2 & 9.58 & 288 & 126 & 9 & 1398 & 1821 & 8179 & 0.07\\
run2 & 5.2 & 8.25 & 1033 & 30 & 10 & 6154 & 7227 & 2773  & 0.004 \\
run3 & 5.2 & -  & 54 & 14 & 12 & 1050 & 1130 & 8870 & 0.01 \\
\hline
\end{tabular}
\caption{Outcomes of the $n$-body simulations. Column~1 is the name of the simulation. Columns~2 and~3 are the orbital radii of Jupiter and Saturn, respectively. Columns~4,~5 and~6 show the number of asteroids that hit the Sun, the Earth, and Jupiter and Saturn. Column~7 shows the number of asteroids that have been ejected.  Column~8 shows the total number of asteroids that have an outcome (either collision or ejection). Column~9 shows the number of asteroids remaining in the simulation. Column~10 shows the probability of an Earth collision for all of the asteroids that have an outcome, $P_{\rm collide}=N_{\rm Earth}/N_{\rm outcome}$.  }
\label{table}
\end{table*}

We have shown how the architecture of the outer giants affects the location and strength of the $\nu_6$ resonance.  However, we have not yet considered how the delivery of asteroids to Earth is affected by the  giant planets being close to the 2:1 MMR.
We used the $n$-body code {\sc Mercury} \citep{Chambers1999} with the Bulirsch-Stoer integrator to model a belt of $10,000$ asteroids in the presence of varying planetary systems.  The asteroid belt was initially the same in all simulations. The particles were distributed uniformly in semi-major axis from $a=1.5\,\rm au$ out to $4.1\,\rm au$. The eccentricity was uniformly distributed in the range $0-0.1$ and the inclination in the range $0-10^\circ$. The longitude of ascending node, argument of perihelion and mean anomaly were all distributed uniformly in the range $0-360^\circ$. We do not expect the results to change qualitatively for a less dynamically excited belt \citep[e.g.][]{Morbidelli1993}.

Each simulation had the Earth at its current location and we considered three different giant planetary systems as described in Table~2. Run1 describes the current solar system, run2 had the giant planets close to the 2:1 MMR and run3 had no Saturn. In all of the simulations the size of the Earth was inflated to $R_{\rm Earth}=10\,\rm R_\oplus$ in order to artificially increase the number of collisions with the Earth. We would expect few to no collisions with the Earth in the simulation if Earth was to be taken with its actual size \citep[see e.g.][]{Smallwood2018}. The inflated Earth increases the number of collisions by a factor of $10-100$ depending upon the relative velocity between the Earth and the asteroid, which varies with semi-major axis of the asteroid. In \cite{Martin2021water} we found  numerically the range to be approximately $20-30$.

Fig.~\ref{splash} shows the outcomes of the simulations as a function of time and the original location of the asteroid.  Asteroids are ejected from the system when their semi-major axis becomes larger than $20\,\rm au$. In Table~\ref{table} we tabulate the outcomes of each simulation and calculate the probability of a collision with the Earth.
Fig.~\ref{splash} shows the initial semi-major axis of asteroids that have an outcome (ejection or collision) and the time of the outcome. The left panel has Jupiter and Saturn at their current locations (run1). Since the asteroids were originally distributed uniformly in semi-major axis, areas of this figure that are white (empty) throughout the simulation, still have stable asteroids remaining at the end of the simulation. We therefore find that with Jupiter and Saturn at their current locations, there are large regions of the asteroid belt that remain stable while some specific resonance locations are cleared out. The outer parts of the belt lie in the chaotic region close to Jupiter, in which there are overlapping MMRs that cause the belt to be unstable.  This configuration satisfies the two conditions described in the Introduction. The large empty areas in this figure show that there is still a significant belt that is able to store asteroids in stable configurations (these stable asteroid orbits do not change significantly during the simulation). The $\nu_6$ resonance provides a high collision probability with the Earth \citep{Ito2006,Martin2021water}. The probability of an Earth collision over the entire simulated belt is 0.07. Of course, this is artificially high because of the inflated size of the Earth. Asteroids from stable regions of the belt may be slowly moved into the $\nu_6$ resonance through effects such as asteroid-asteroid interactions, gas drag, the Yarkovsky effect \citep{Farinella1998} or implantation from farther out \citep{Jewitt2014,Raymond2017}.

The right hand panel of Fig.~\ref{splash} shows the simulation without Saturn (run3). In this case, the inner parts of the asteroid belt are much more stable than when Saturn is included (as in run1) since there is no $\nu_6$ resonance. There are very few collisions or ejections in the inner belt. The few outcomes  that occur are a result of MMRs with Jupiter. This configuration satisfies the first criterion in the Introduction - there is a stable asteroid belt. However, there is no efficient mechanism of delivering the asteroids to Earth. The probability of an Earth collision in run3 is 7 times lower than in the simulation with Saturn at its current location (run1). Note that \cite{Smallwood2018} considered impact rates with a varying location for Saturn. Provided that the $\nu_6$ resonance fell within the belt, the rate of impacts remained high. 

The middle panel of Fig.~\ref{splash} shows the outcomes in the simulation in which the giant planets are in resonance (run2). The giant planets display chaotic but stable behaviour \citep[see also][]{Izidoro2016}.  This leads to stochastic jumps of  the secular resonances and a rapid excitation of the asteroid orbits. This simulation, in contrast to the other simulations, shows that most of the asteroid belt is unstable. The resonance between the giant planets leaves stable asteroids only inside of about $1.6\,\rm au$. The timescale for the ejection of the asteroids is very short. Thus, there is no stable belt in a system with the giant planets in resonance and we do not satisfy the first criterion described in the Introduction.  Despite the unstable belt, there are very few asteroid collisions with the Earth. In fact, the probability of an Earth collision is smaller in this case by a factor 2.5 compared to run3 and by a factor of 18 compared to run1.

\section{Discussion and Conclusions}

Assuming that asteroid impacts are indeed necessary for the emergence of life, we suggest that there are two requirements on an asteroid belt for a habitable exoplanet to actually be inhabited. First, the exoplanetary system must have the equivalent of an asteroid belt, which may require a giant planet to form outside of the snow line radius with a low eccentricity. Second, the asteroid belt must have a mechanism to deliver asteroids from stable regions of the belt.  MMRs with the inner giant planet may not provide a large enough supply of asteroids and therefore a secular resonance, such as the $\nu_6$ resonance in the solar system, may have to fall within the asteroid belt region. This suggests that a second giant planet is required, and its location must fall within a relatively narrow radial region in order for sufficient asteroids to collide with the habitable planet and its eccentricity must be low. This requires a certain degree of  fine tuning and it makes the solar system somewhat special \citep[see also][]{Livio2020}.   Still, current observations of pairs of giant exoplanets suggest that this configuration is not uncommon.

It is interesting to examine two exoplanetary systems that are high on the list of candidates for the search for life on other exoplanets that are around M-dwarfs. TRAPPIST-1 has 7 exoplanets including three Earth-like planets within the habitable zone \citep{Gillon2016}.  While, there is no evidence for a belt in this system \citep{Marino2015}, we do note that our own asteroid belt in an exoplanetary system would not be observable \citep[e.g.][]{Wyatt2007}. Comet impacts from a Kuiper belt equivalent in the system could occur through perturbations from an external perturber \citep{Kral2018,Raymond2021}.  Proxima Centauri has two detected exoplanets, one of which is close to Earth's size and within the habitable zone \citep{AngladaEscude2016}. It is not certain whether an asteroid belt exists in this system \citep[e.g.][]{Siraj2020}.

We caution that our conclusions are based on the classical picture of solar system formation. Recently it has been suggested that planetesimals may form in rings that we see in the observations of DSHARP disks \citep{Andrews2018}.  Further, it has been suggested that the formation of the planets in the solar system was through rings of planetestimals that formed through pressure bumps \citep{Morbidelli2021,Izidoro2021,Carrera2021}. In this picture, asteroid belts lie between the rings and no giant planet is required for their formation. However, if we require a secular resonance to lie within an asteroid belt, then we still need two giant planets in the system. Furthermore, we do not take into account the migration histories of the giant planets that may deplete the asteroid belt through the motion of the resonances \citep{Minton2011,Clement2020}.

\begin{acknowledgements}

We thank an anonymous referee for a thorough review and providing useful comments. This research has made use of the NASA Exoplanet Archive, which is operated by the California Institute of Technology, under contract with the National Aeronautics and Space Administration under the Exoplanet Exploration Program  \citep[DOI 10.26133/NEA12,][]{doi}. Computer support was provided by UNLV’s National Supercomputing Center.  RGM acknowledges support from NASA through grant 80NSSC21K0395.

\end{acknowledgements}




\bibliographystyle{aasjournal}
\bibliography{martin} 



\end{document}